\documentclass[12pt]{article}
\usepackage{graphicx,amsmath,amssymb,euscript,psfrag,epsf,epsfig}

\usepackage{a4wide}
\newcommand{\LQCD}{\Lambda_\mathrm{QCD}}

\newcommand{\be}{\begin{equation}}
\newcommand{\ee}{\end{equation}}
\newcommand{\bea}{\begin{eqnarray}}
\newcommand{\eea}{\end{eqnarray}}
\newcommand{\nl}{\nonumber \\}

\newcommand{\as}{\alpha_s}

\newcommand\Star[3]{\biggl(\frac{#1}{#2}\biggr)_*^{\left[{#3}\right]}}

\def\slash#1{#1 \hskip-0.45em /}
\def\Slash#1{#1 \hskip-0.59em /}

\def\beq{\begin{eqnarray}}
\def\eeq{\end{eqnarray}}

\def\eps{\epsilon}

\def\be{\begin{equation}}
\def\ee{\end{equation}}

\def\np{n_+}
\def\nm{n_-}

\def\mc{m_c}
\def\mp{m_p}
\def\Li2{\text{Li}_2}
\def\Im{\text{Im}}

\def\nslash{\rlap{\hspace{0.02cm}/}{n}}

\def\pslash{\rlap{\hspace{0.02cm}/}{p}}
\def\vslash{\rlap{\hspace{0.02cm}/}{v}}

\def\LQCD{\Lambda_{\rm QCD}}

\def\L{{\cal L}}

\begin{document}

\begin{titlepage}

\begin{flushright}
SI-HEP-2005-18 \\[0.2cm]
December 13, 2005\\
revised: April 13, 2006
\end{flushright}

\vspace{1.2cm}

\begin{center}
\Large\bf\boldmath
Can $\bar B\to X_c \ell \bar\nu_\ell$ help us extract $|V_{ub}|$?
\unboldmath 
 \end{center}

\vspace{0.5cm}

\begin{center}
{\sc H.~Boos, Th.~Feldmann, T.~Mannel, B.D.~Pecjak} \\[0.1cm]
{\sf Theoretische Physik 1, Fachbereich Physik,
Universit\"at Siegen\\ D-57068 Siegen, Germany}
\end{center}

\vspace{0.8cm}
\begin{abstract}
\vspace{0.2cm}\noindent
We study radiative corrections to 
$\bar{B}\to X_c \ell \bar{\nu}_\ell$  
decays assuming the power counting 
$m_c\sim \sqrt{\LQCD m_b}$ for the charm-quark mass. 
Concentrating on the shape-function
region, we use effective field-theory methods to calculate the 
hadronic tensor at NLO accuracy.
From this we deduce a shape-function independent relation
between partially integrated $\bar B \to X_c\ell\bar\nu_\ell$ and
$\bar B \to X_u\ell\bar\nu_\ell$ spectra to leading power in $1/m_b$, 
including first-order corrections in the strong coupling constant. 
This may provide an independent cross-check on the determination of 
the CKM element $|V_{ub}|$.

\end{abstract}


\end{titlepage}

\section{Introduction}

A central goal of the $B$ physics program is to accurately
determine the CKM parameter $|V_{ub}|$.  A complication
is that experiments cannot measure the total rate for
inclusive $\bar B\to X_u \ell \bar\nu_\ell$ decays, because 
part of the available phase space is dominated 
by a much larger background from  
$\bar B\to X_c \ell \bar\nu_\ell$ decays.
In fact, data for inclusive $b\to u$ transitions
is available only in the shape-function region, where the
final-state hadronic jet carries a large energy on the 
order of the $b$-quark mass $m_b$, but a relatively small
invariant mass squared on the order of $\LQCD m_b$.  

The study of inclusive $B$ decays in 
the shape-function region using soft-collinear 
effective theory (SCET) has received much 
attention in recent years 
\cite{Bauer:2003pi,Bosch:2004th,powercorrections,Neubert:2004dd,Lee:2005pk}.  
Predictions for decay
distributions are available in the form of
factorization formulas  
which separate the physics from the three scales 
$m_b \gg \sqrt{\LQCD m_b} \gg \LQCD$.  At leading
order in $1/m_b$, the factorization formula
takes the form
\be\label{eq:fact}
H\cdot J\otimes S.
\ee
The hard function $H$ and the jet function $J$ 
are perturbatively calculable functions 
depending on quantities at the hard scale $m_b$ 
and the hard-collinear (jet) scale $\sqrt{\LQCD m_b}$ respectively.
The shape function $S$ is a non-perturbative 
function defined in terms of a
non-local HQET matrix element \cite{earlyshape}.  
There are two basic 
strategies for reducing shape-function related 
hadronic uncertainties in the measurement of $|V_{ub}|$. 
The first is to extract the shape function in one 
process and use it as input for
other processes, the second is to construct shape-function 
independent relations between different decay distributions. 
Common implementations of these strategies use 
$\bar B\to X_s\gamma$ in 
combination with $\bar B\to X_u \ell \bar\nu_\ell$ decay spectra 
\cite{Neubert:1993um,MannelRecksiegel,
Leibovich:1999xf, phenom, Hoang:2005pj,Lange:2005qn,Lange:2005xz}.

Assuming the power counting $m_c^2\sim \LQCD m_b$ for the
charm-quark mass, parts of the phase space for 
$\bar B\to X_c \ell \bar\nu_\ell$ decays  lie in the 
shape-function region \cite{MannelNeubert,MannelTackmann}.
Work performed in \cite{Boos:2005by} showed that
the singly differential spectrum in a 
certain kinematic variable has much in common 
with the $P_+$ spectrum in $\bar B\to X_u \ell \bar\nu_\ell$ decays.  
In particular, at tree level and excluding
power corrections,
this spectrum is directly proportional to the 
leading-order shape function.  This raised the possibility of using 
data from inclusive decays into charm quarks to learn about the 
leading-order shape function.  The analysis 
in  \cite{Boos:2005by} concentrated on the classification
of sub-leading effects in the $\LQCD/m_b$ expansion 
at tree level, while the question of 
radiative corrections was left open.

In this paper we calculate the perturbative corrections
to $\bar B\to X_c \ell \bar\nu_\ell$ decays in the shape-function
region. We show that our one-loop result for the hadronic 
tensor can be written in the factorized form (\ref{eq:fact}).  
Moreover, the hard function $H$ and the shape function $S$ are identical 
for inclusive $b\to u$ and $b\to c$ transitions;
the charm-quark mass affects only  the jet function $J$. 
This allows us to construct a simple, 
shape-function independent relation between the 
$\bar B\to X_c \ell \bar\nu_\ell$ and
$\bar B\to X_u \ell \bar\nu_\ell$ decay spectra, 
which may provide an independent cross-check for
the determination of $|V_{ub}|$.

The paper is organized as follows.  In Section~\ref{sec:SCET}
we discuss some aspects of SCET needed in our analysis.
We use this to calculate the hadronic tensor at one loop 
in Section~\ref{sec:fact}.  In Section~\ref{sec:uspec}
we present results for the partially integrated spectrum
needed in our phenomenological discussion and examine
some issues related to the definition of the charm-quark
mass.  A relation between partially integrated
$b\to u$ and $b\to c$ spectra is derived and 
studied in Section~\ref{sec:phenom}.  We conclude
in Section~\ref{sec:conclusions}.


\section{SCET  for  $\bar B \to  X_c \ell \bar\nu_\ell$ transitions}
\label{sec:SCET}
In this section we review some aspects of SCET
\cite{Bauer:2000ew,Bauer:2000yr,Bauer:2001yt,Beneke:2002ph}
needed to describe inclusive 
$b\to c$ transitions in the shape-function region.
The effective theory facilitates the separation of scales
and sets up a systematic expansion in the small parameter
\be
\lambda^2\sim \frac{m_c^2}{m_b^2} \sim \frac{\LQCD}{m_b}.
\ee
At the level of Feynman diagrams, this separation 
of scales is achieved by evaluating QCD
integrals using the method of regions  
\cite{Beneke:1997zp}, and the construction of SCET
is closely related to this diagrammatic analysis.  
To apply this method one first identifies the
momentum regions which give rise to leading-order
on-shell singularities in loop diagrams.  The integrand
is expanded in $\lambda$ as appropriate for the particular
region before performing the integral.  
Once all the regions  are identified,  
their sum is equal to the full theory integral, up to
higher-order terms in $\lambda$.

Applying the method of regions to 
inclusive $b\to u$ transitions in
the  shape-function region, where
the jet momentum and the jet energy satisfy
$p^2\sim m_b^2 \lambda^2$ and 
$E\sim m_b$, one finds contributions 
from  hard, hard-collinear, and soft 
regions. SCET is constructed in such a way that
the hard-collinear and soft regions are 
contained in effective theory
fields and operators, while the hard
region is contained in Wilson coefficients multiplying
these operators.  For the $b\to c$ transitions dealt with 
in this paper, we will always work in the kinematic
region where the jet momentum and the jet energy satisfy
$p^2-m_c^2 \sim p^2 \sim  m_b^2 \lambda^2$ and 
$E\sim m_b$. It is apparent that the set 
of regions is identical to that in the charmless case; 
one must replace $p^2\to p^2-m_c^2$ in the hard-collinear
propagators, but the $\lambda$
expansion, and thus the regions calculation, works the same.  
Therefore, the relevant version of SCET is very
similar to that for charmless decays.  
The objects of interest are the SCET Lagrangian and currents, 
which we now discuss in turn.


\subsection{SCET Lagrangian and mass renormalization}

\label{subsec:lagrangian}

The leading-order SCET Lagrangian for a hard-collinear quark 
with mass $m_c$ interacting with soft and hard-collinear gluons
is  (see for instance \cite{Rothstein:2003wh,Leibovich:2003jd,Boos:2005by})
\bea\label{eq:L}
\L&=& \bar\xi
\left(i\nm D + (i \Slash{D}_{\perp \rm hc}-m_c)\frac{1}{i\np D_{\rm hc}}
(i \Slash{D}_{\perp \rm hc}+m_c)\right)\frac{\slash{n}_+}{2}\xi
+ \L_{{\rm YM}}.
\eea
Here $\xi$ is a hard-collinear quark field, and the covariant
derivatives are defined as 
$i n_- D = in_-\partial + g n_- A_{\rm hc} + g n_- A_s$
and  $i D_{\rm hc} = i \partial + g A_{\rm hc}$.
We have introduced two light-like vectors $n_{\pm}$, which
satisfy $\np \nm =2$.
The Yang-Mills Lagrangian $\L_{{\rm YM}}$ for the
soft (hard-collinear) sector is the same as in QCD, but restricted
to soft (hard-collinear) fields. 

In massless SCET, the Lagrangian is not 
renormalized, in the sense that no new operators or 
non-trivial Wilson coefficients are induced by radiative
corrections.  The reasoning for this was given
in \cite{Beneke:2002ph}, and involves showing
that certain momentum regions give rise to scaleless integrals.
These arguments also apply to the SCET Lagrangian
(\ref{eq:L}), because the $\lambda$
expansion is unaffected by the presence of a 
quark mass in the hard-collinear propagators, 
as we have emphasized above. 

On the other hand, mass renormalization plays a non-trivial
role in our analysis, and will be needed in the 
next section when we calculate the one-loop jet function.
We pause here to discuss mass renormalization in SCET.
Later on we will study the differential
spectrum in the variable 
$$u=\nm p -m_c^2/\np p \,,$$
where $m_c$ may be taken as the pole mass
(in the massless case $u$ reduces to the variable
$p_+ = \nm p$). 
In Section~\ref{sec:change}, we will discuss alternative mass definitions which
induce a change in the jet function of order $\alpha_s$.
 
Mass renormalization in SCET is closely related to the usual
QCD prescription, which follows from the observation that 
the self-energy diagram in SCET can be obtained from
the $\lambda$ expansion of the corresponding QCD diagram.  
This has been pointed out for the massless case
in \cite{Bauer:2000ew}, and for the massive case
with $m^2 \ll \LQCD m_b$ in \cite{Chay:2005ck}.
We have confirmed 
that it also holds for the case $m_c^2 \sim \LQCD m_b$.  
In full QCD, the one-loop fermion propagator is
\be\label{eq:QCDprop}
G(p)= \frac{i}{\pslash - m_c - \Sigma (p)},
\ee
where the fermion self-energy reads
\be
\Sigma(p)=\pslash \, \Sigma_V (p^2) + m_c \, \Sigma_S (p^2).
\ee
Analogously, the one-loop fermion propagator in SCET is 
\be\label{eq:SCETprop}
G_\xi(p)= \frac{i}{u - \Sigma_\xi (u,\np p)}\frac{\nslash_-}{2} \ .
\ee 
For simplicity we consider a frame where $p_\perp =0$, 
such that $u=\nm p-m_c^2/\np p$. We obtain the SCET fermion 
self-energy $\Sigma_\xi (u, \np p)$ by expanding the 
QCD propagator (\ref{eq:QCDprop}) to leading order 
in $\lambda$ and 
matching it with the SCET propagator (\ref{eq:SCETprop}), 
which gives the result 
\be
\Sigma_\xi(u, \np p) = u\, \Sigma_V(p^2) + \frac{m_c^2}{\np p}\, 2
 \left(\Sigma_V(p^2) + \Sigma_S(p^2)\right). 
\ee
Taking into account mass renormalization, 
the renormalized fermion propagator in SCET is
\be\label{SCETpropren}
\hat G_\xi(p)= \frac{i}{u - \Sigma_\xi (u,\np p) 
- \frac{\delta (m_c^2)}{\np p}}\frac{\nslash_-}{2},
\ee 
where $\delta(m_c^2)= 2 m_c \delta m_c$.
The propagator has a pole for 
$p^2=m_c^2 \Leftrightarrow u=0$, from which we get 
\be\begin{split}
\frac{\delta(m_c^2)}{\np p}&=-\Sigma_\xi (0,\np p)
=-\frac{m_c^2}{\np p}\, 2 
\left(\Sigma_V (m_c^2) + \Sigma_S(m_c^2)\right)\\  
&= - 6 \frac{\mc^2}{\np p} \frac{ C_F\as}{4\pi}
\biggl(\frac1\eps - \ln \biggl(\frac{\mc^2}{\mu^2}\biggr) 
+ \frac43\biggr) 
\end{split}
\label{eq:dmpol}
\ee
as the corresponding mass counterterm in the pole scheme.

\subsection{SCET transition current}\label{subsec:current}

Unlike the Lagrangian, the SCET representation of the
weak transition current involves non-trivial 
hard matching coefficients. The  matching onto SCET
takes the form \cite{Bauer:2000yr,Beneke:2002ph}
\bea
e^{im_b v x}\bar c(x)\gamma^\mu(1-\gamma_5) b(x)&\to& 
\sum_{i=1}^3 \int ds \, \tilde C_i(s,m_b) 
(\bar\xi W)(x+s\np)\Gamma_i^\mu\, h_v(x_-)
\nonumber 
\\
&=& \sum_{i=1}^3 \, C_i(\np p, m_b) 
(\bar\xi W)(x)\Gamma_i^\mu \, h_v(x_-)
\label{eq:current}
\eea
where $h_v$ is the heavy-quark field defined in HQET,
$W$ is a hard-collinear Wilson line,
and $p$ is the momentum of the hard-collinear quark.  
The Dirac structures are chosen as
\be
\Gamma_i^\mu=\{\gamma^{\mu}(1-\gamma_5),\,v^\mu(1+\gamma_5),
\,\frac{\nm^\mu}{\nm v}(1+\gamma_5)\}.
\ee

One calculates the hard coefficients $C_i$ by matching
the one-loop corrections to the current from QCD
to SCET. The QCD diagrams receive contributions from
hard, hard-collinear and soft momentum regions.  Since SCET
is constructed to reproduce the results for the 
hard-collinear and soft regions, it is only 
the hard region of the QCD diagrams 
which contributes to the matching
conditions. However,  the Taylor-expanded integrand
for the hard region does not depend on the hard-collinear
scale $m_c^2$, so the matching conditions are the 
same as in the massless case. We can therefore read off
the result for the coefficients $C_i$ from \cite{Bauer:2000yr}.
The matching conditions also involve a current renormalization
factor, which accounts for the divergent part of the hard
diagrams.  Its explicit form is \cite{Bauer:2000yr}
\be\label{eq:ZJ}
Z_J=1+ \frac{C_F \as}{4\pi}\left(-\frac{1}{\eps^2}
+\frac{2}{\eps}\ln\frac{\np p}{\mu}
-\frac{5}{2\eps}\right).
\ee
We will need this renormalization factor in our calculation
of the hadronic tensor in the next section.


\section{Hadronic tensor at one loop}
\label{sec:fact}

In this section we calculate the one-loop
corrections to the hadronic tensor for 
$\bar B\to X_c \ell \bar \nu_\ell$ decays in the 
shape-function region, always working to
leading order in $\lambda$. 
The hadronic tensor contains all 
the QCD effects in the semi-leptonic
decay and is the starting point for deriving 
differential decay distributions. 
We define the hadronic tensor as
\be
W^{\mu\nu}=\frac{1}{\pi}{\rm Im}\langle \bar B(v)|
T^{\mu\nu}|\bar B(v)\rangle ,
\ee
where we use the state normalization 
$\langle \bar B(v)|\bar B(v)\rangle=1$. 
The current correlator 
$T^{\mu\nu}$ is given by 
\be\label{eq:correlator}
T^{\mu\nu}=i \int d^4 x e^{-i q\cdot x}{\rm T}
\{J^{\dagger \mu}(x) J^{\nu}(0)\},
\ee
where $J^\mu=\bar c \gamma^\mu(1-\gamma_5)b$ 
is the flavor-changing weak transition 
current discussed above.

The one-loop result for the hadronic tensor 
can be written in the factorized form 
\be\label{eq:fact2}
W^{\mu\nu}=\sum_{i,j=1}^3  \frac 12 
\text{tr}\biggl(\bar\Gamma^{\mu}_j\frac{\nslash_-}{2}
\Gamma^{\nu}_i \frac{1+\vslash}{2}\biggr)
H_{ij}(\np p)\int d\omega J(u-\omega,\np p) S(\omega), 
\ee
where $p\equiv m_b v-q$ is the jet 
momentum in the parton
model. The hard functions $H_{ij}$, the jet 
function $J$, and the shape function 
$S$ contain physics at the scales 
$m_b^2$, $\LQCD m_b$, and $\LQCD^2$, respectively. 
The limits of integration in the convolution 
integral are determined by the facts that the 
shape function has support for 
$-\bar\Lambda \leq \omega <\infty$ and the 
jet function has support for $u-\omega\geq 0$.

The procedure leading to (\ref{eq:fact2}) is familiar 
from charmless decay  and involves
a two-step matching procedure \cite{Bauer:2003pi,Bosch:2004th}.  
In the first step, one
integrates out hard fluctuations at the scale $m_b$ 
by matching
the hadronic tensor calculated in QCD onto that
calculated in SCET. The associated matching coefficients are
the hard functions $H_{ij}$. Since these coefficients take 
into account the hard region of the QCD diagrams, and this 
region is unaffected by the presence of a quark mass in
the hard-collinear propagators,
they are identical to those in the massless
case. One finds $H_{ij}=C_j C_i$, where
the $C_i$ are the hard Wilson coefficients defined
in (\ref{eq:current}).
In the second step, one integrates out hard-collinear
fluctuations at the scale $\LQCD m_b $
by matching the hadronic tensor calculated
in SCET onto that calculated in HQET. 
The matching coefficient from this step is the jet
function $J$.
This function is obviously more
complicated than in massless SCET, since it 
can depend on $m_c^2$ as well as $p^2$. We will 
calculate it in the following subsection. 
However, the final low-energy theory is still HQET,
and the matrix element defining the shape function
is the same as in charmless decays. For this 
reason, we can write our result in the form (\ref{eq:fact2}).

\subsection{One-loop jet function}

\begin{figure}[!t]
\begin{center}
\includegraphics[width=1\textwidth]{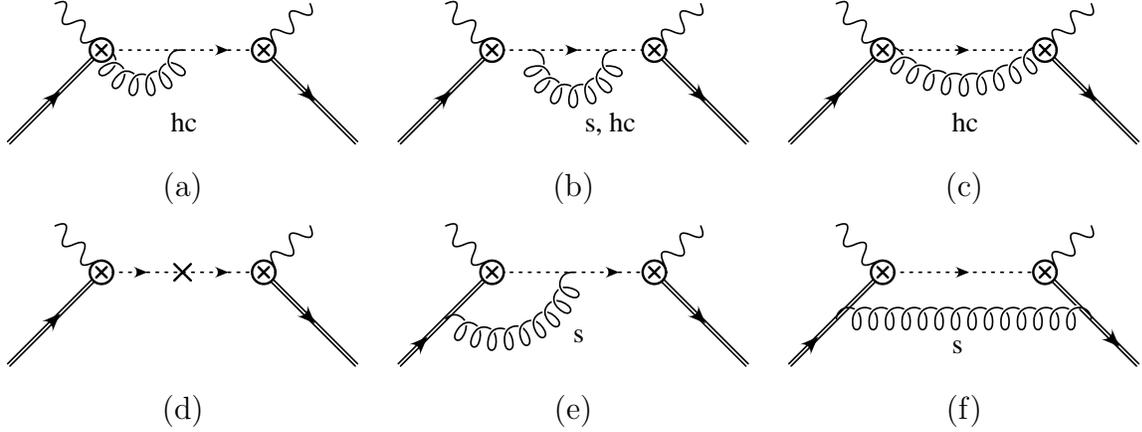} 
\parbox{0.94\textwidth}{
\caption{\label{fig:correlator} \small
The one-loop SCET graphs contributing
to the current correlator.  Mirror graphs are not shown. Graph (d)
shows the insertion of a counterterm from mass renormalization.}
}
\end{center}
\end{figure}

The calculation of the one-loop
jet function is conceptually identical to 
that for the massless case 
\cite{Bauer:2003pi,Bosch:2004th}, and we will
closely follow the treatment in \cite{Bosch:2004th}.
The jet function is the matching coefficient
between the hadronic tensor calculated in SCET and that
calculated in HQET. The relevant SCET diagrams are shown in 
Fig.\ref{fig:correlator}.  We calculate them
in the parton model, using 
on-shell heavy quark states carrying a residual 
momentum $k$ satisfying $vk=0$.  
We work with dimensional regularization in 
$d=4-2\eps$ dimensions, using the Feynman gauge. 
The result for the graphs involving hard-collinear
gluon exchange, including
the counterterm from mass renormalization in the pole scheme
(\ref{eq:dmpol}), 
can be written as 
\be
D_{hc}^{(1)}=\mathcal{J}_{hc}^{(1)} 
\,\left[\bar h_v \, \bar\Gamma_j^\mu \, \frac{\slash{n}_-}{2}\, 
\Gamma_i^\nu\, 
h_v\right],
\ee
where
\bea
\mathcal{J}_{hc}^{(1)}& =& 
\frac{C_F\as }{4\pi}\frac{i}{u'} \Bigg\{
\frac{4}{{\eps}^2} + \frac{3}{\eps} - 
\frac{4}{\eps}\ln \left( \frac{-\np p\,u'}{\mu^2} \right)\nl
&+&
7 - \frac{{\pi }^2}{3} - 3\,
\ln \biggl(\frac{-\np p\,u'}{\mu^2} \biggr) + 
2\,\ln^2 \biggl( \frac{-\np p \,u'}{\mu^2} \biggr) 
\nl
&+ & 
\frac{2\,{\pi }^2}{3} 
- 4\,\Li2\biggl(1+\frac{\mp }{u'}\biggr)
\nl
&+&
\frac{\mp}{\mp + u'} 
- \frac{\mp\,\left( \mp + 2\,u' \right)} 
{{\left( \mp + u' \right) }^2} 
\,\ln \biggl( -\frac{u'}{\mp} \biggr)
\Bigg\}.
\label{eq:Dhc}
\eea
Here $u^\prime = u+\nm k$, $\as\equiv\as(\mu)$, 
and $\mp \equiv \mc^2/\np p$. We have checked that
our result (\ref{eq:Dhc}) agrees with the corresponding result 
in \cite{Chay:2005ck} when expanded
in $m_c^2/p^2$ and translated to the $\overline{\rm MS}$ scheme.
The graphs involving soft gluon 
exchange give
\be
D_{s}^{(1)}=\mathcal{J}_{s}^{(1)}\,
\left[\bar h_v \, \bar\Gamma_j^\mu \, \frac{\slash{n}_-}{2}\, 
\Gamma_i^\nu\, 
h_v\right],
\ee
where
\be
\mathcal{J}_{s}^{(1)} =
\frac{C_F\as }{4\pi}\frac{i}{u'} \Bigg\{
-\frac{2}{\eps^2}+\frac{2}{\eps}+
\frac{4}{\eps}\ln\left(\frac{-{u^\prime}}{\mu}\right)-
\frac{3\pi^2}{2}-4\ln\left(\frac{-{u^\prime}}{\mu}\right)
-4\ln^2\left(\frac{-{u^\prime}}{\mu}\right)\Bigg\}.
\ee
The sum of the $1/\eps$ poles in 
$\mathcal{J}_{hc}^{(1)}+\mathcal{J}_{s}^{(1)}$ is removed
by current renormalization in SCET, which is implemented
by applying a factor of $Z_J^2$ (see \ref{eq:ZJ}) 
to the bare current correlator.
This renormalization factor
is related to the divergent part of the hard region of the QCD 
diagrams, which was integrated out in the first 
step of matching.  That it cancels the $1/\eps$ poles from the SCET
diagrams, which are due to both hard-collinear and
soft regions, shows that we have indeed constructed the 
appropriate version of SCET.  Moreover,
the pole structure for each individual region is the 
same as in the massless
case. It follows that the hard and shape functions
obey the same renormalization group evolution as in the
massless case, a fact which we will use when discussing
decay distributions in the next section.

We can interpret the imaginary part of the
finite pieces of the SCET diagrams
as one-loop corrections to the factorized expression  
(\ref{eq:fact2}). They take the form 
$J^{(0)}\otimes S_{part}^{(1)}+J^{(1)}\otimes S_{part}^{(0)}$,
where the superscript $(n)$ denotes the $n$-loop correction
to each function, and the $\otimes$ stands for a convolution.
The tree-level functions are  $J^{(0)}=\delta(u-\omega)$ and
$S^{(0)}_{part}=\delta(\omega+\nm k)$. As in the massless
case, the one-loop correction to the shape function 
in the parton model is related to $\mathcal{J}_{s}^{(1)}$.  
To show this, we take its imaginary part, 
which can be expressed in terms of 
star distributions, defined as 
\cite{DeFazio:1999sv}
\bea
\int_{\leq 0}^M du \, F(u)\left(\frac{1}{u}\right)_*^{[m]}&=&
\int_0^M du\frac{F(u)-F(0)}{u}+F(0)\ln\biggl(\frac{M}{m}\biggr), 
\eea
\bea
\int_{\leq 0}^M du \, F(u)
\left(\frac{\ln(u/m)}{u}\right)_*^{[m]}&=&
\int_0^M du\frac{F(u)-F(0)}{u}\ln\frac{u}{m}+\frac{F(0)}{2}
\ln^2\biggl(\frac{M}{m}\biggr).
\eea
It is not difficult to derive the following formulas
\bea\label{eq:star1}
&&-\frac{1}{\pi}{\rm Im} 
\left[\ln\left(-\frac{u}{m}\right)\frac{1}{u}\right]=
\left(\frac{1}{u}\right)_*^{[m]}\nl
&&-\frac{1}{\pi}{\rm Im} \left[
\ln^2\left(-\frac{u}{m}\right)\frac{1}{u}\right]
=2\left(\frac{\ln (u/m)}{u}\right)_*^{[m]}
-\frac{\pi^2}{3}\delta(u),
\eea
where to take the correct branch of the logarithms we 
reinstored $u\equiv u+i \eps$. We then obtain
\bea
&&J^{(0)}\otimes S^{(1)}_{part}=S^{(1)}_{part}(u^\prime)
=\frac 1\pi \Im 
\biggl[i\,\mathcal{J}_{s, finite}^{(1)} \biggr] \nl
&&=-\frac{C_F \as}{4\pi}\left[\frac{\pi^2}{6}\delta(u^\prime)
+4\Star{1}{u'}{\mu}+8\Star{\ln(u'\,/\mu)}{u'}{\mu}\right],
\label{eq:Spart1}
\eea
which is identical to the  
one-loop result calculated in HQET.
The jet function is related to the imaginary
part of the finite piece of $\mathcal{J}_{hc}^{(1)}$
\cite{Bosch:2004th}.  
In particular, we have
\be
J^{(1)}\otimes S^{(0)}_{part}=J^{(1)}(u^\prime,\np p)=\frac 1\pi \Im 
\biggl[i\,\mathcal{J}_{hc, finite}^{(1)}\biggr].
\ee
Using (\ref{eq:star1}) along with
\be
-\frac 1\pi \Im \biggl[\frac{1}{u'}
\biggl[\Li2\biggl(1+\frac{m}{u'}\biggr)\biggr]\biggr]= 
-\Star{\ln(u'\, /m)}{u'}{m}+\frac{1}{u'} 
\, \ln\biggl(1+\frac{u'}{m}\biggr)\,\theta(u'),
\ee
we find that the  
jet function to $\mathcal{O}(\alpha_s)$ is given by
\bea\label{eq:jet}
J(u',\np p)=\delta (u')&+&\frac{C_F\as }{4\pi}\biggl\{
\left( 7 - \pi^2 \right) \delta(u')  
- 3 \Star{1}{u'}{\mu^2/\np p}
  + 4 \Star{\ln(u'\, \np p/\mu^2)}{u'}{\mu^2/\np p}\nl
&&+\biggl(\frac{u'}{(\mp+u')^2} 
- \frac{4}{u'} \ln\biggl(1+\frac{u'}{\mp}\biggr)
\biggr)\,\theta (u') +\left( 1 + \frac{{2\pi }^2}{3} 
\right) \,\delta(u')\nl
&&-\Star{1}{u'}{\mp} + 4\,
\Star{\ln(u'\, /\mp)}{u'}{\mp}\biggr\}.
\eea
The first line of (\ref{eq:jet}) reduces to the
result for the massless case in the limit
$m_c\to 0$ \cite{Bauer:2003pi,Bosch:2004th}, while
the second and third lines are unique to decay into 
charm quarks, and vanish for  $m_c \to 0$.

We have also compared our calculation with the one-loop OPE
result for $b \to c \ell \nu$ in 
\cite{Aquila:2005hq}.\footnote{For the comparison with earlier
calculations in \cite{Trott:2004xc,Uraltsev:2004in} see the
detailed discussion in \cite{Aquila:2005hq}.}
For this purpose we re-expand the factorized expression
for the hadronic tensor (\ref{eq:fact2}) in $\alpha_s$. 
Using the notation of \cite{Bosch:2004th}, 
the component $W_1$ of the hadronic tensor, for instance, 
can be written as
\beq
  \frac{W_1}{2} &=& \frac{1}{n_+p}
 \left[ H_{\rm 11} \, \delta(u') 
  + J^{(1)}(u',n_+p) + S^{(1)}_{\rm part}(u') \right] 
  + {\cal O}(\alpha_s^2) \, ,
\label{eq:W1}
\eeq
where the soft and jet contribution are given in
(\ref{eq:Spart1}) and (\ref{eq:jet}), 
and the hard contribution reads 
\cite{Bosch:2004th}
\beq
  H_{\rm 11} &=& 1 + \frac{\alpha_s C_F}{4\pi}
 \left(
 -4 L^2 + 10 L - 4 \ln y - \frac{2 \ln y}{1-y} - 4 {\rm Li}_2(1-y) -
 \frac{\pi^2}{6}-12 \right)
\eeq
with $y = n_+p/m_b$ and $L= \ln \left[y m_b/\mu\right]$. This has 
to be compared with the corresponding expression in
\cite{Aquila:2005hq} 
\beq
W_3^{\protect \cite{Aquila:2005hq}} &=&
\pi\, m_b^2 \, W_1 \, ,
\eeq
where the result for $W_3^{\protect \cite{Aquila:2005hq}}$ has to
be expanded, using the SCET power counting
$m_c^2/m_b^2 \sim \LQCD/m_b \ll 1$. After some tedious,
but straight-forward manipulations, we indeed find agreement with
(\ref{eq:W1}). The comparison for the remaining components $W_{4,5}$ of
the hadronic tensor is much easier because at order $\alpha_s$
they receive contributions from the hard functions only. 
Therefore the limit $m_c/m_b \to 0$ in the corresponding expressions
in \cite{Aquila:2005hq} can be performed directly to
recover the results for  $W_{4,5}$ from \cite{Bosch:2004th}.

\section{The partially integrated $U$ spectrum}
\label{sec:uspec}

\def\omhat{\hat \omega}

From the results of the previous section we can derive
any differential decay distribution. We focus here on
the $U$ spectrum, because of its relation to the 
$P_+$ spectrum from charmless decays. 
We pointed out in the last section that the
hard and shape functions are unaffected by the presence
of the charm-quark mass, and thus obey the same
renormalization group equations as  in the charmless 
case. In the remainder of the paper, we will work 
with the renormalization-group 
improved formulas derived in \cite{Bosch:2004th}.  
After integrating over the lepton energy and neglecting higher-order
terms in $\lambda$, 
the doubly differential spectrum in the variables 
$u$ and $y$
is given by
\be
  \frac{1}{\Gamma_c} \, \frac{d^2\Gamma_c}{du \, dy}
=
  e^{V_H(m_b,\mu_i)} \int_{-\bar \Lambda}^u  d\omega \,
  y^{2-a} (6 - 4 y)\,{\cal H}(y) \, J(u-\omega, m_b y,\mu_i) 
\, S(\omega,\mu_i)\,,
\label{eq:theo}
\ee
where $-\bar\Lambda \leq u \leq y m_b - m_c^2/y m_b$ and
$m_c/m_b \leq y \leq 1$. 
Note that after resummation all of the functions 
are to be evaluated at the intermediate scale 
$\mu_i\sim m_c$. We have introduced
the renormalization-group factors   
\be
 a = \frac{16}{25} \ln \frac{\alpha_s(\mu_i)}{\alpha_s(m_b)},
\ee
and $V_H(m_b,\mu_i)$, which resum logarithms between
the hard and the jet scale. 
The exact form of $V_H$ 
can be found in \cite{Bosch:2004th},
and the hard function ${\cal H}$
can be derived from the functions $H_{ij}$ 
in the same reference.
The total $b\to c$ rate to order $\as(m_b)$ 
in the OPE is given by \cite{Nir:1989rm} 
\be
\Gamma_c= |V_{cb}|^2\left(\frac{G_F^2 m_b^5}{192\pi^3}\right)
\left[f\left(\frac{m_c^2}{m_b^2}\right)
+ \frac{C_F\alpha_s(m_b)}{4\pi}
\left(\frac{25}{2}-2\pi^2\right) g\left(\frac{m_c^2}{m_b^2}\right)\right].
\ee
At leading order in $\lambda$ we can set the phase-space
factors $f,g$ to unity, although higher-order corrections
may be important numerically, as we shall discuss 
in Section \ref{sec:kincorr}.

\begin{figure}[!t]
\begin{center}
\includegraphics[width=0.47\textwidth]{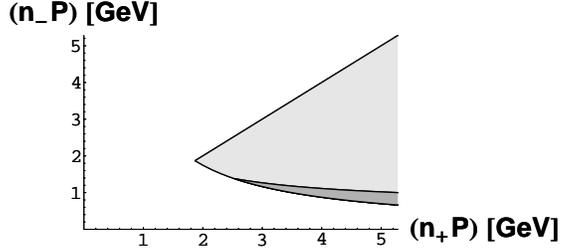} 
\parbox{0.94\textwidth}{
\caption{\label{fig:phasespace}\small 
Illustration of the shape-function region singled out by
the cut on the variable $U$. The light-grey region shows the
physical phase space $M_D^2/n_+P \leq n_-P \leq n_+P \leq M_B$.
The dark-grey part shows the shape-function region with
$\Delta=0.65$~GeV and $m_c = 1.36$~GeV.
}}
\end{center}
\end{figure}

It is useful to change from partonic to hadronic variables and 
define\footnote{Notice that $\omega$ 
is defined with the opposite sign in \cite{Bosch:2004th}, 
whereas our convention for $\hat S(\omhat)$
coincides.}
\be
  U = u + \bar \Lambda, \qquad
  \omhat = \omega + \bar \Lambda, \qquad
  \hat S(\omhat) = S(\omega) \ .
\ee
The relation between the hadronic momenta
$P^\mu$ and the partonic momenta $p^\mu$
is given by  $n_\pm P = n_\pm p + \bar\Lambda$,
where $\bar \Lambda = M_B - m_b$.
This leads to
\beq
  U 
  &=& n_-P - \frac{m_c^2}{n_+P} + {\cal O}(\lambda^4).
\eeq
To stay in the shape-function region, we need to restrict
the phase-space integration to values of $U \sim \lambda^2 m_b$.
In order to preserve a close correspondence with the treatment
of the $\bar B \to X_u \ell \bar\nu_\ell$ spectrum 
in \cite{Bosch:2004th}, we introduce a cut $U < \Delta$, 
with $\Delta$ being around 600~MeV.
The effect of this cut on the physical phase space
in the variables $P_-=n_+P$ and $P_+=n_-P$ is illustrated in
Fig.~\ref{fig:phasespace} for typical values $\Delta=0.65$~GeV
and $m_c = 1.36$~GeV.
The fraction of events with $U <\Delta$ is then given by
\beq
F_c(\Delta) &= & \frac{\Gamma_c(U < \Delta)}{\Gamma_c}
\nonumber \\[0.2em]
&=& 
e^{V_H}
\int_0^\Delta d\omhat 
\int_{ \frac{m_c}{m_b}}^1 dy
\int_0^{\Delta} dU \, y^{2-a}(6-4 y)\,
{\cal H}(y) \, J(U-\omhat,\, y m_b) 
\, \hat S(\omhat).
\label{eq:Fcmaster2}
\eeq
A short calculation shows that the lower limit
of the integration over $y$ can be set to zero, up to terms of  order  
$(m_c/m_b)^{3-a}$. 
After making this simplification, the integration
limits are identical to those in $b\to u$ decays.
In fact, the integrals over the $\as$ corrections
from the hard function ${\cal H}$ and 
the first line of (\ref{eq:jet}) are identical
to the charmless case.  We will give
explicit results below. 

The new terms  
relevant to decay into charm quarks
are contained in the last two lines of (\ref{eq:jet}).  
After integration over $U$ the result for these terms is 
\be\begin{split}
e^{V_H}\int_0^\Delta d\hat\omega \hat S(\hat\omega) 
\int_{0}^1 dy\,& y^{2-a}(6-4 y)
\frac{C_F\alpha_s(\mu_i)}{4\pi}\bigg\{
\frac{2\pi^2}{3}-\ln(y\Delta_{\omhat} )
+2 \ln^2( y \Delta_{\omhat} )\\&
+\frac{1 }{1+ y \, \Delta_{\omhat} }
+\ln(1+ y \Delta_{\omhat} )+4\Li2(-y \Delta_{\omhat} )\bigg\},
\end{split}
\ee
where $\Delta_{\omhat}=(\Delta-\omhat)m_b/m_c^2$.
The integrals over $y$ can be evaluated in terms of 
the master integrals
\beq
G_1(n,x) &=& \int_0^1  dy \, 
\frac{y^{n}}{1+ x y}
   = \frac{  {}_2 F_1(1,n+1;n+2;-x)}{n+1} , \\
G_2(n,x)&=& \int_0^1 dy \,y^{n}\ln(1+x y)=\frac{1}{n+1}
\left(\ln(1+x )-\frac{1}{n+1}\right)+ \frac{G_1(n,x)}{n+1} , \\
G_3(n,x)&=&\int_0^1 dy\,y^{n} \Li2(- x y)=
\frac{ \Li2(- x)}{n+1}+ \frac{G_2(n,x)}{n+1},
\eeq
where the hypergeometric function $ {}_2 F_1$ 
has a series expansion
\be
{}_2 F_1(a_1,a_2;a_3;z)=\sum_{k=0}^\infty 
\frac{(a_1)_k\, (a_2)_k}{(a_3)_k}\frac{ z^k}{k!},
\qquad (a_i)_k=\frac{\Gamma(a_i+k)}{\Gamma(a_i)}.
\ee
To express the final results in a compact way,
we introduce
\bea
g_n(a,X)&=&\frac{6G_n(2-a,X)-4G_n(3-a,X)}{T(a)} \ ,
\eea
and make use of the functions defined in \cite{Bosch:2004th}
\bea
f_2(a)&=&-\frac{30-12a + a^2}{(6-a)(4-a)(3-a)}, \qquad
f_3(a)=\frac{2(138-90a+18a^2-a^3)}{(6-a)(4-a)^2(3-a)^2},\nl
T(a)&=& \frac{2(6-a)}{(4-a)(3-a)} \ .
\eea
The final result can be written as $F_c=F_u+F_m$. The fraction
$F_u$ is the result for $b\to u$ decays
\bea
F_u(\Delta)&=& T(a)e^{V_H(m_b,\mu_i)}
\int_0^\Delta d\omhat \,\hat S(\omhat,\mu_i) 
\, f_u\left(\frac{m_b (\Delta-\omhat)}{\mu_i^2}\right) \ , 
\label{eq:Fu} \\[0.3em]
f_u(x) &=&
1+
\frac{C_F\alpha_s(m_b)}{4\pi}H(a) \nl
&&+\frac{C_F\alpha_s(\mu_i)}{4\pi}\left[
2\ln^2 x +
\big(4f_2(a)-3\big)\ln x
+\big(7-\pi^2-3f_2(a)+2f_3(a)\big)\right].
\nonumber
\eea
An expression for $H(a)$ can 
be found in \cite{Bosch:2004th}.
The fraction $F_m$ is an additional piece unique to decay
into charm quarks, which vanishes when $m_c\to 0$. In the pole scheme,
it is given by
\bea\label{eq:extra}
F_m(\Delta)&=&T(a) \, e^{V_H(m_b,\mu_i)} 
\int_0^\Delta d\omhat \,\hat S(\omhat,\mu_i) \, 
f_m\left(\frac{m_b(\Delta-\omhat)}{m_c^2}\right)
\ , \\[0.3em]
f_m(x) &=& \frac{C_F\alpha_s(\mu_i)}{4\pi} \bigg[
2 \ln^2 x +
\big(4 f_2(a)-1\big) \ln x 
\nl
&&
+\frac{2\pi^2}{3}-f_2(a)+2 f_3(a)+
 g_1(a,x)+g_2 (a,x)+
4g_3(a,x)\bigg] 
\ .
\nonumber 
\eea
From the partially integrated spectrum $F_c(\Delta)$ we can
obtain the corresponding $U$ spectrum by differentiation, which
results in
\beq
  \frac{1}{\Gamma_c} \frac{d\Gamma_c}{dU}
&=& T(a) \, e^{V_H(m_b,\mu_i)} 
\int_0^U d\omhat \left( \frac{d}{d\omhat} \hat S(\omhat,\mu_i) \right) 
  \left[f_u\left(\frac{m_b(U-\omhat)}{\mu_i^2}\right)
     + f_m\left(\frac{m_b(U-\omhat)}{m_c^2}\right) \right]
\cr &&
\label{eq:dGammac}
\eeq
where we have used integration by parts and $\hat S(0)=0$.

\subsection{Change of charm-mass definition}

\label{sec:change}

So far, our analysis has been performed with $m_c$ 
defined in the pole scheme. 
The effect of changing the charm-mass definition
according to
$$
 m_c \to \tilde m_c = m_c - \delta m \ ,
$$
where $\delta m \sim m_c \alpha_s(\mu_i)$, is two-fold. First, the
input value for the charm-quark mass in $F_m(\Delta)$ is changed. Since
the explicit charm-mass dependence in $F_m$ is already an ${\cal
O}(\alpha_s)$ correction, this effect is formally of
order $\alpha_s^2$. 
Second, the jet function receives a perturbative correction 
proportional to $\delta m$. It can be obtained from 
the tree-level jet function by taking into account the appropriate
shift in the spectral variable,
\beq
  \delta(u-\omega) 
    &\simeq&
 \delta\left(\tilde u - \omega\right)
         - \frac{2 \tilde m_c \delta m}{n_+p} \, \delta'(\tilde u-\omega), 
\eeq
where $\tilde u = n_-p - \tilde m_c^2/n_+p$ is defined using the
new mass definition. Inserting the extra term into (\ref{eq:Fcmaster2}),
one obtains an additional contribution to $F_c(\Delta)$,
\beq
  F_c(\Delta) & \to & 
  F_c(\Delta) - e^{V_H}\int_0^\Delta d\omhat \int_0^1 dy
\int_0^{\Delta}dU \, y^{2-a}(6-4 y)\,
 \frac{2 \tilde m_c \delta m}{y m_b} \, \delta'(U-\omhat) \, 
\hat S(\omhat) \nonumber \\[0.2em]
 &=& F_c(\Delta) - e^{V_H} \, T(a+1) 
     \, \frac{2 \tilde m_c \delta m}{m_b} \, \hat S(\Delta) \ .
\eeq
In order to see the scheme-independence of physical observables
to a fixed order in $\alpha_s$, one has to keep in mind
that the relation between the hadronic momenta and the spectral
variable $U$ is also changed.
Therefore, the result for $F_c(\Delta)$ in two different mass schemes 
should be compared at two different values of the
cut-off parameter $\Delta$, 
\beq
  \tilde U = n_-P - \frac{\tilde m_c^2}{n_+P} \ < \ \tilde \Delta
     \simeq \Delta + \frac{2 \tilde m_c \delta m}{n_+P} \ .
\eeq
such that $\tilde F_c(\tilde \Delta)$ 
in the new scheme reads
\beq
 \tilde F_c(\tilde \Delta) &=& F_u(\tilde \Delta) 
    + F_{\tilde m}(\tilde \Delta) - e^{V_H} \, T(a+1) 
     \, \frac{2 \tilde m_c \delta m}{m_b} \, \hat S(\tilde \Delta) \ .
\label{eq:Ftilde}
\eeq
Expanding the upper limit $\tilde \Delta$ around $\Delta$ in the
leading-order term in $F_u(\tilde \Delta)$ and neglecting terms of
order $\alpha_s^2$, we explicitly find the scheme-independence of
our result,
\beq
  \tilde F_c(\tilde \Delta) &=& F_c(\Delta) + {\cal O}(\alpha_s^2) \ .
\eeq
Still, the convergence of the perturbative series 
at a given value of $\Delta$  might be rather different for 
different mass definitions. In addition to the pole scheme, we will consider 
two further examples, namely
\begin{itemize}
  \item the potential-subtracted (PS) scheme, where \cite{Beneke:1998rk}
        \beq
          m_c^{\rm PS}(\mu_f) &=& m_c -
             \frac{C_F \alpha_s(\mu_i)}{\pi} \, \mu_f + {\cal O}(\alpha_s^2)
        \eeq
        with $\mu_f \simeq 1$~GeV,
  \item the $\overline{\rm MS}$ scheme, where
        \beq
          \bar m_c(\mu_i) &=&  m_c \left[
           1 + \frac{C_F \alpha_s(\mu_i)}{4\pi} \left(
               3 \ln \frac{m_c^2}{\mu_i^2} - 4 \right) + {\cal O}(\alpha_s^2)
         \right]\ .
        \eeq
\end{itemize}

\subsection{Numerical predictions}

In this section we study the numerical predictions for $F_c(\Delta)$,
taking into account mass-scheme and shape-function dependence.
We start by summarizing the parameter values used in the subsequent analysis.
The hard scale is fixed to the $b$\/-quark mass, $m_b = 4.65$~MeV.
The default value for the intermediate (jet) scale is $\mu_i = 1.5$~GeV.
We use the PS scheme as our default mass scheme, taking
$m_c^{\rm PS}(\mu_f=1~{\rm GeV})=1.36 
$~GeV.
The charm-quark pole mass is taken as $1.65$~GeV,
and the $\overline{\rm MS}$ mass at the jet scale as
$\bar m_c(\mu_i)=1.20$~GeV.
We use 2-loop running for $\alpha_s$ 
with $\LQCD^{(n_f=4)}=345$~MeV, corresponding to 
$\alpha_s(m_b)=0.22$ and $\alpha_s(\mu_i)=0.37$.

\begin{figure}[!t]
\begin{center}
\includegraphics[width=0.47\textwidth]{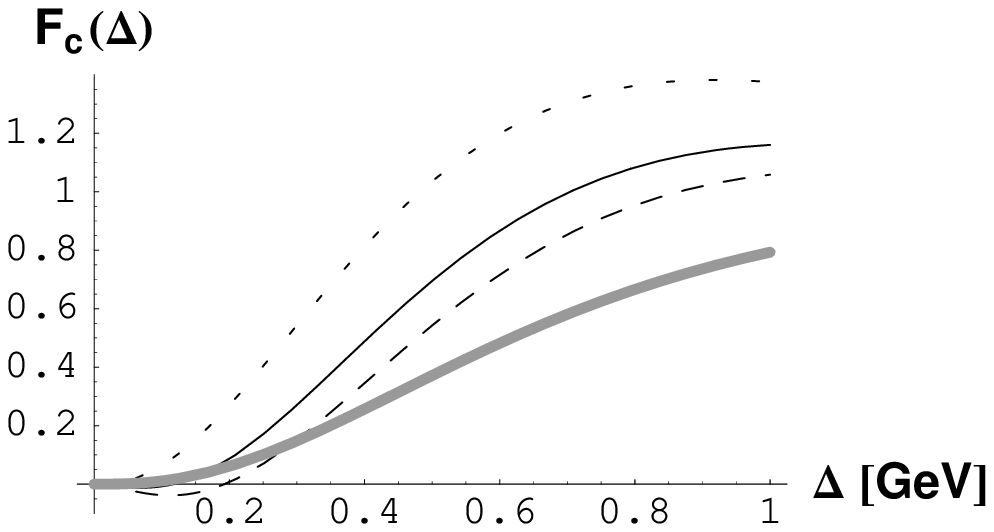} 
\ \ \ \ \
\includegraphics[width=0.47\textwidth]{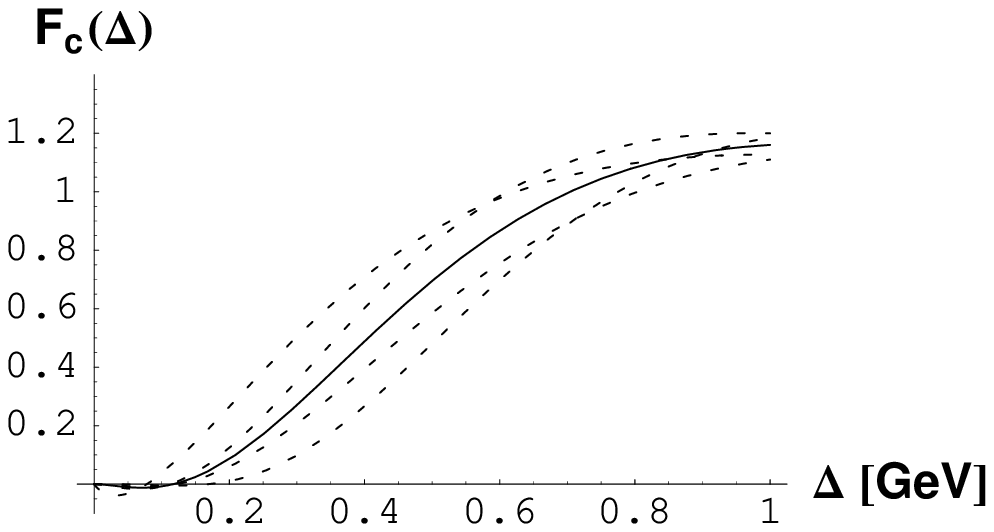} 
\parbox{0.94\textwidth}{
\caption{\label{fig:Fc}\small 
Predictions for partially integrated spectra in
inclusive semi-leptonic $b \to c $~decays: 
Left: NLO prediction for $F_c(\Delta)$ using the
the default scenario S5 \cite{Bosch:2004th} in the PS scheme 
(solid line) compared to  
the $\overline{\rm MS}$ scheme (long-dashed line)
and the pole scheme (short-dashed line). Also plotted is 
the LO result (thick grey line).
Right:  NLO prediction for $F_c(\Delta)$ using
the default scenario S5  in the PS scheme 
(solid line) compared to scenarios S1, S3, S7, S9 
(dashed lines).
}}
\end{center}
\end{figure}

For the numerical estimate we have to specify a model for the
shape function, which we take from \cite{Bosch:2004th}\footnote{
{ 
Notice that we have chopped off the radiative tail in 
$\hat S(\hat\omega)$, which does not contribute for the value of
$\Delta$ that we are considering.
}
}:
\beq
  \hat S(\hat\omega,\mu_i) &=& \frac{1}{\Lambda}
  \left[1-\frac{C_F \alpha_s(\mu_i)}{4\pi}
    \left(\frac{\pi^2}{6}-1\right)\right]
    \, \frac{b^b}{\Gamma(b)}
    \left(\frac{\hat \omega}{\Lambda}\right)^{b-1}
    \exp\left(-b \frac{\hat\omega}{\Lambda} \right) \ .
\label{eq:S5}
\eeq
We use $\Lambda=0.685$~GeV and
$b = 2.93$ as our default (scenario ``S5'' in \cite{Bosch:2004th}). 
In Fig.~\ref{fig:Fc} we compare the results
for different mass schemes and different input shape functions as
a function of the cut-off $\Delta$.
The following observations can be made:
\begin{itemize}
  \item For values of $\Delta \sim 600$~MeV, the NLO 
        corrections are large and positive. 

  \item Above some critical value $\Delta_{\rm max}$, the NLO corrections
        become so large that
        the fraction $F_c$ exceeds 1, and 
        therefore our result should not be trusted anymore.

  \item The critical value $\Delta_{\rm max}$ 
        amounts to about $480$~MeV in the pole scheme, $700$~MeV
        in the PS scheme, and $860$~MeV in the $\overline{\rm MS}$
        scheme.

  \item The model dependence from the input shape function 
        amounts to an uncertainty of about 25\%.
  
\end{itemize}

\subsection{Power corrections}

\begin{figure}[!t]
\begin{center}
\includegraphics[width=0.5\textwidth]{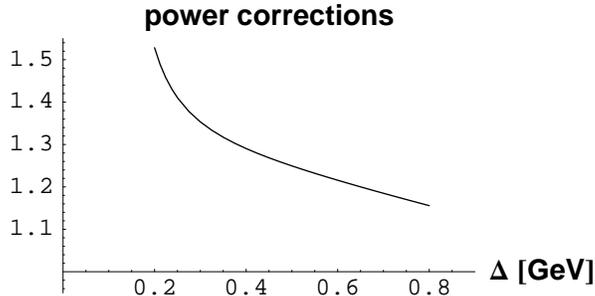} 
\parbox{0.94\textwidth}{\caption{\label{fig:ps} \small
Effect of kinematic power corrections proportional to
$\hat S(\omhat)$: The curve shows the NLO prediction for 
$F_c(\Delta)$ including the power corrections to the tree-level
result in (\ref{eq:ps}), normalized to the leading-power result
(\ref{eq:Fu},\ref{eq:extra}).}}
\end{center}
\end{figure}

\label{sec:kincorr}

Our NLO calculation has been restricted to leading power in the
$1/m_b$ expansion. Power corrections arise from two sources.
First, one encounters new non-perturbative structure in the form of
sub-leading shape functions.
Second, there are kinematic power corrections proportional to
$m_c^2/m_b^2 \sim \lambda^2$ and $u \sim \lambda^2 m_b$. The
phase-space integration leads to logarithms $\ln (m_c^2/m_b^2)$
which can numerically enhance some of the power-suppressed terms.\footnote{
These phase-space logarithms can be resummed using renormalization-group
techniques, see \cite{Bauer:1996ma}.}
Whereas the
estimate of sub-leading shape function effects is model dependent,
the kinematic corrections multiplying the leading-order
shape function can be calculated explicitly. 
We shall do this at tree level only, where we can
use the results of \cite{Boos:2005by} to find
\beq
    \frac{1}{\Gamma_c} \, \frac{d\Gamma_c}{dU}
&=&
  \left\{ 1 - 
  \frac{U-\bar\Lambda}{m_b} \left( \frac{14}{3} + \frac{m_c^2}{m_b^2} 
     \left(\frac{215}{6} + 3 \,
    \ln\frac{m_c^2}{m_b^2} \right) \right) + 
     {\cal O}(u^2,\lambda^5)  \right\}
 \hat S(U) 
\nonumber \\[0.3em] && {} + \mbox{sub-leading shape functions} \ .
\label{eq:ps}
\eeq
The omitted terms are negligible in
the portion of phase-space we are interested in. 
The numerical effect of the power corrections in (\ref{eq:ps}) is
plotted in Fig.~\ref{fig:ps}. We see that $F_c(\Delta)$ is enhanced
by about 20\% at $\Delta=0.65$~GeV. Since we cannot control the remaining
power corrections from sub-leading shape functions in a model-independent
way, we consider this number as a rough estimate for the magnitude of the 
systematic uncertainties
associated with power corrections.


\section{Relating $b \to c$ and $b \to u$ decays}\label{sec:phenom}

For the extraction of the CKM parameter $|V_{ub}|$
one would like to have a shape-function-independent relation 
between the $\bar B \to X_u \ell \bar\nu_{\ell}$ and 
$\bar B \to X_c \ell \bar\nu_{\ell}$ decay
spectra.   
In what follows we focus on a relationship between
the $P_+$ spectrum in $b\to u$ decays and the $U$ spectrum
in $b\to c$ decays.
This relation can be obtained in a similar way
as discussed for the comparison of $\bar B \to X_s \gamma$ 
and $\bar B \to X_u \ell \bar\nu_{\ell}$
in \cite{Lange:2005qn}.  In the present case, this involves 
constructing a weight function $W$ such that 
\be\label{eq:weight}
\int_0^\Delta dP_+ \,\frac{d\Gamma_u}{d P_+}=
\frac{\Gamma_u}{\Gamma_c}  \int_0^\Delta dU
\,W(\Delta,U) \,\frac{d\Gamma_c}{dU}
\simeq \frac{|V_{ub}|^2}{|V_{cb}|^2}  \int_0^\Delta dU
\,W(\Delta,U) \,\frac{d\Gamma_c}{dU}
,
\ee
where we have used that $\Gamma_u/\Gamma_c = |V_{ub}/V_{cb}|^2$
to leading power in $\lambda$.
By measuring the  partial decay rate $\Gamma_u(P_+<\Delta)$ 
in $b\to u$ decays, as well as the $d\Gamma_c/dU$ spectrum 
in $b \to c$ decays, we can determine $|V_{ub}|$.
The theoretical input is the weight function $W$,
which we can calculate from the results in
(\ref{eq:extra},\ref{eq:Ftilde}):
\bea
W(\Delta, U) &=&  1- f_m\left(\frac{m_b(\Delta - U)}{(m_c^{\rm PS})^2}\right)
+\frac{C_F\alpha_s(\mu_i)}{4\pi}
 \frac{T(a+1)}{T(a)} \, \frac{8 \mu_f m_c^{\rm PS}}{m_b} \, 
\delta(\Delta -U)
\nl &&
{} + {\cal O}(\alpha_s^2) + \mbox{power corrections},
\label{eq:weightfunc}
\eea
in the PS scheme. We do not attempt to 
include power corrections here, but must be aware 
that they add a systematic uncertainty of at least
20\%.

At the moment, we do not have explicit experimental information
on the $U$~spectrum in $\bar B \to X_c\ell\bar \nu_\ell$, 
so to illustrate how our method works we will have to rely on
some theoretical input. In the following subsections we
will consider two approaches: In the first we will use the
theoretical prediction (\ref{eq:dGammac}) for the $b \to c$ spectrum to
obtain the $b\to u$ spectrum from the weight-function analysis
with (\ref{eq:weightfunc}). In the second approach, we will construct a
simple toy spectrum which takes into account possible
charm-resonance effects.

\subsection{Numerical analysis using theoretical $b \to c$ spectrum}

\begin{figure}[!t]
\begin{center}
\includegraphics[width=0.45\textwidth]{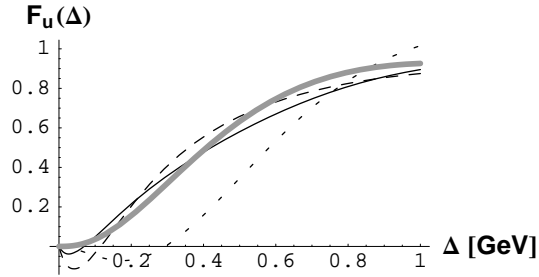}
\parbox{0.94\textwidth}{
\caption{\label{fig:weightvsdir1}\small
Predictions for $F_u(\Delta)$ using the weight function~(\ref{eq:weightfunc})
and the theoretical $b \to c$ spectrum~(\ref{eq:dGammac}) on the basis
of the shape-function model S5 in (\ref{eq:S5}).
Solid line: PS~scheme. Long-dashed line: $\overline{\rm MS}$~scheme.
Short-dashed line: Pole scheme. For comparison, we also show the
direct computation of $F_u(\Delta)$ from (\ref{eq:Fu})
(thick grey line).}
}
\end{center}
\end{figure}

\begin{figure}[!t]
\begin{center}
\includegraphics[width=0.45\textwidth]{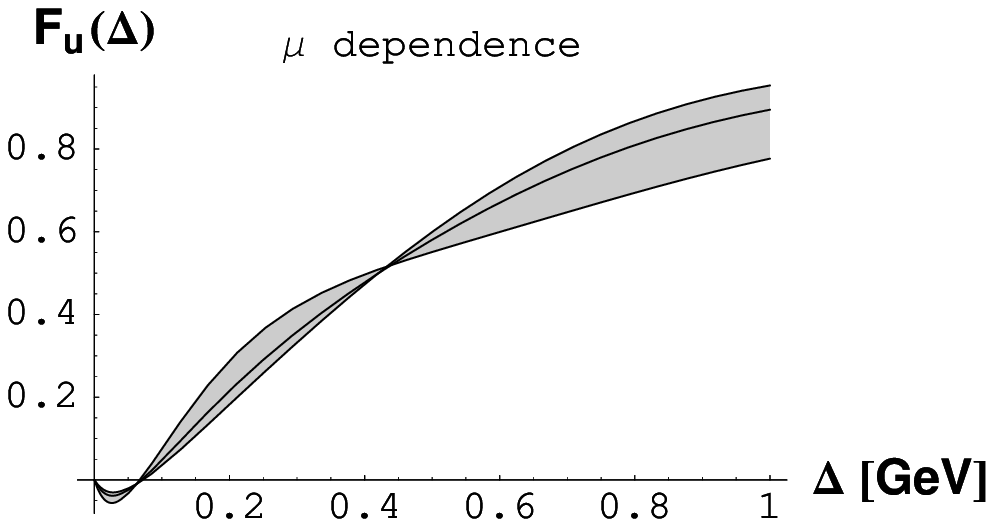}
\hspace{1.3em}
\includegraphics[width=0.45\textwidth]{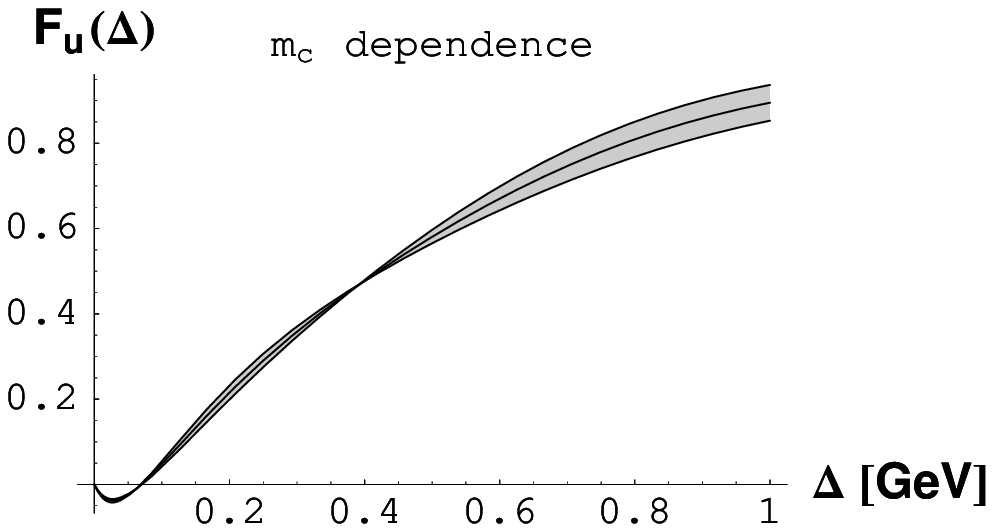}
\parbox{0.94\textwidth}{
\caption{\label{fig:weightvsdir2}\small
Predictions for $F_u(\Delta)$ using the weight function~(\ref{eq:weightfunc})
and the theoretical $b \to c$ spectrum~(\ref{eq:dGammac}) on the basis
of the shape-function model S5 in (\ref{eq:S5}). The left plot shows the
scale dependence ($1$~GeV$< \mu_i < 2.25$~GeV)
of the result using the PS scheme. The right plot shows the
effect of varying $m_c^{\rm PS}$ by $\pm 0.15$~GeV around its
default value.
}
}
\end{center}
\end{figure}

In this sub-section we carry out the weight-function analysis using
the theoretical $b\to c$ spectrum from (\ref{eq:dGammac}) as input.
This will help us estimate some of the perturbative uncertainties
inherent to our approach. 
We start by constructing the partially integrated $b\to u$ 
spectrum $F_u(\Delta)$ from the theoretical $b \to c$
spectrum (\ref{eq:dGammac})
and the weight function $W(\Delta,U)$, 
using the shape-function model from (\ref{eq:S5}).  
In  Fig.~\ref{fig:weightvsdir1} we compare the 
so-obtained results for $F_u(\Delta)$ in
the PS, pole, and $ \overline{\rm{MS}}$ schemes. We also show the result
of the direct computation from (\ref{eq:Fu}).  
The difference between the curves is
formally an ${\cal O}(\alpha_s^2)$ effect, and thus gives a rough 
measure of higher-order perturbative effects. We observe 
that in the pole scheme this effect is quite large for values 
of $\Delta$ below $700$~MeV or so.
At our reference point $\Delta =650$~MeV we obtain
\beq
  F_u(0.65~{\rm GeV}) &=& 0.71 \qquad \mbox{from (\ref{eq:dGammac}) and 
     (\ref{eq:weight}), PS scheme} \ ,
\cr
  F_u(0.65~{\rm GeV}) &=& 0.62 \qquad \mbox{from (\ref{eq:dGammac}) and 
     (\ref{eq:weight}), pole scheme} \ ,
\cr
  F_u(0.65~{\rm GeV}) &=& 0.76 \qquad \mbox{from (\ref{eq:dGammac}) and 
     (\ref{eq:weight}), $\overline{\rm MS}$ scheme} \ ,
\label{eq:Futheo}
\eeq
compared to
\beq
  F_u(0.65~{\rm GeV}) &=& 0.79 \qquad \mbox{from (\ref{eq:Fu})} \ ,
\eeq
from which we deduce a residual scheme dependence for $F_u(\Delta)$
of about 10-15\%.

In Fig.~\ref{fig:weightvsdir2} we investigate the explicit $\mu_i$\/
and $m_c$ dependence induced by the weight function
(to isolate these effects, we fix $m_c$ and $\mu_i$ in 
the theoretical expression (\ref{eq:dGammac}) for the $b \to c$
spectrum). The charm-mass dependence is a small effect, 
less than 10\% for reasonably large values of $\Delta$.
The dependence on the factorization scale $\mu_i$ is still sizeable, 
about 10-15\%. The perturbative uncertainties related to
the scheme and factorization-scale dependence 
could be resolved by calculating the $\alpha_s^2$ 
corrections to the jet function in the massive case.

\subsection{Numerical analysis using a toy spectrum}

The purpose 
of this sub-section is to point out some aspects of the
weight-function analysis that would be important when dealing
with the physical $b\to c$ spectrum.
A distinctive feature of this spectrum is that 
that the lowest-lying spin-symmetry doublet of 
charmed states $D$ and $D^*$ 
already makes up about 80\% of the semi-leptonic rate.
For an on-shell $D(D^*)$ meson we formally have
\be
U_{D(D^*)} = \frac{M_{D(D^*)}^2-m_c^2}{n_+P} \ll \Delta 
\qquad (n_+P \sim m_b) \ .
\ee
Therefore, about 80\% of the $U$ spectrum is centered around ``small'' 
values of $U$ (we put ``small'' in quotation marks, because numerically
$(M_D^2-m_c^2)/m_b \simeq 350~$MeV in the PS scheme).


We will perform the weight-function analysis on
a toy spectrum which takes this resonance structure into account. 
We construct this spectrum by assuming that the doubly-differential
decay spectrum is concentrated along the $D/D^*$\/-pole 
and modulated by some function $f(y)$,
\beq
  \frac{1}{\Gamma_c} \, \frac{d^2\Gamma_c}{d(n_-P) dy} &\simeq &
  \frac{m_b \, y^2}{\overline M_D^2} \, f(y) \, 
  \delta\left(y - \frac{\overline M_D^2}{n_-P \, m_b}\right) \,
\label{model2}
\eeq
where $\int_0^1 dy \, f(y) =1$, and $\overline M_D =1.975$~GeV is a 
weighted average of the $D$ and $D^*$ masses.  
We can derive the $U$ spectrum (in a given mass scheme) 
from this model by taking
$$
  n_-P = U + \frac{m_c^2}{y m_b}
$$
and performing the integral over $y$, which yields
\beq
   \frac{1}{\Gamma_c} \, \frac{d\Gamma_c}{dU} &\simeq &
  \frac{{\overline M}_D^2-m_c^2}{m_b U^2} \, 
  f\left(\frac{{\overline M}_D^2-m_c^2}{m_b U}\right) \, 
  \theta\left(U - \frac{\overline M_D^2-m_c^2}{m_b}\right) \,
\label{model1} .
\eeq
For the following discussion, we use a simple parameterization
\beq
  f(y) &=& \frac{\Gamma(2+\alpha+\beta)}{\Gamma(1+\alpha)\Gamma(1+\beta)}
  \, y^\alpha \, (1-y)^\beta
\eeq
and fix the parameters $\alpha = 3.66$ and $\beta=-0.51$ 
by requiring that $F_c(\Delta)$ at $\Delta=650$~MeV,
and $d\Gamma_c/dU$ at $U=550$~MeV coincide
with the theoretical expressions in the PS scheme.\footnote{
The reference point for $F_c(\Delta)$ is sufficiently below the
critical value $\Delta_{\rm max}=700$~MeV, and
that for $d\Gamma_c/dU$ is sufficiently above 
the exclusive threshold $U_{\rm min}\simeq 450$~MeV.} 

\begin{figure}[!t]
\begin{center}
\includegraphics[width=0.45\textwidth]{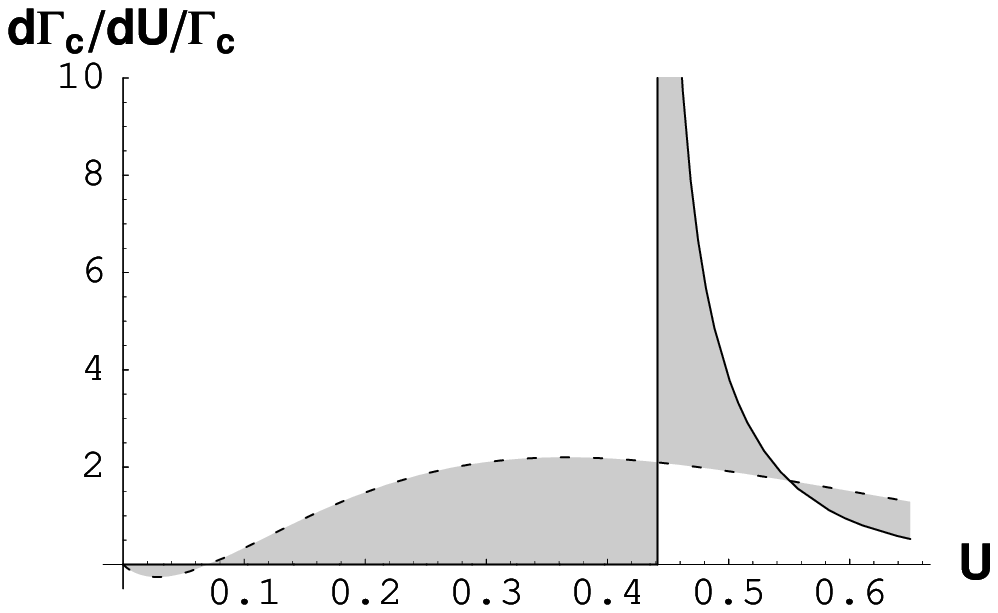}
\hspace{1.3em}
\includegraphics[width=0.45\textwidth]{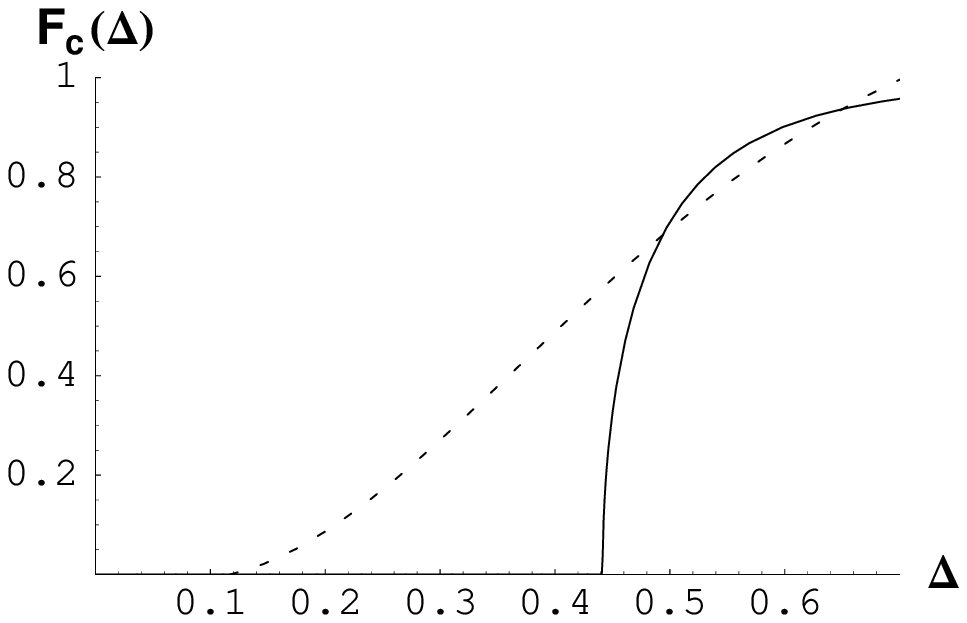}
\parbox{0.94\textwidth}{
\caption{\label{fig:model}\small
The $U$ spectrum in $\bar B \to X_c \ell \bar \nu_\ell$ decays:
Theoretical prediction using the default model S5 for the shape 
function (dashed line) vs.\ phenomenological model 
(solid line) assuming the dominance of a single $D$\/-meson pole. The model 
parameters are adjusted to reproduce the value of the spectrum
(left) at $U=550$~MeV as well as the integrated spectrum (right)
at $\Delta=650$~MeV.}
}
\end{center}
\end{figure}

\begin{figure}[!t]
\begin{center}
\includegraphics[width=0.45\textwidth]{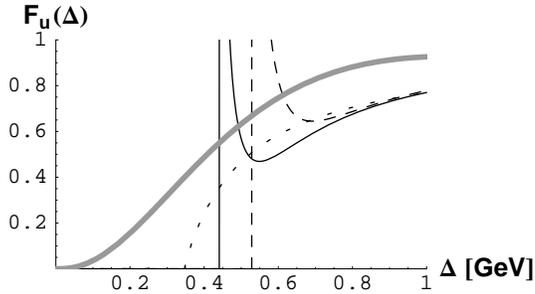}
\parbox{0.94\textwidth}{
\caption{\label{fig:schemedep} \small 
Predictions for the partial rate $F_u(\Delta)$ in
$\bar B \to X_u \ell \bar \nu_\ell$ from a toy spectrum in 
$\bar B \to X_c \ell \bar \nu_\ell$ using the NLO weight function.
Comparison of PS scheme (solid line), pole scheme (dotted),
and $\overline{\rm MS}$ scheme (dashed)
with the theoretical result (\ref{eq:Fu}) using scenario S5
(thick grey line).}}
\end{center}
\end{figure}

In Fig.~\ref{fig:model} we compare the  $U$ spectrum 
from our toy model with the theoretical prediction in the
PS scheme, using the shape-function model (\ref{eq:S5}) as
input. Fig.~\ref{fig:schemedep} shows predictions 
for $F_u(\Delta)$ obtained by applying the weight-function analysis
to our toy spectrum, as well as the theoretical 
curve obtained from (\ref{eq:Fu}).
We see that at smaller values of $\Delta$ the sensitivity
to the resonance structure and dependence on the mass scheme
is sizeable. On the other hand, for larger
values of $\Delta$ the resonance structure is washed out, and
the predictions obtained in different mass schemes converge.
The sensitivity to the resonance structure at moderate values of 
$\Delta$ means that the phenomenologically acceptable
window for $\Delta$ in the shape-function approach is smaller 
than in the $b\to u$ case, where the contributions 
from the charmless ground states 
$\pi$, $\eta$, $\rho$ and $\omega$ add up to only 
about 25\% of the total semi-leptonic $b \to u$ rate 
and are moreover centered at values of $P_+$ not much larger 
than 100~MeV.  

These observations have important
implications for extracting $|V_{ub}|$ by relating 
partially integrated $b\to u$ and $b\to c$ decay spectra.
To apply our results to $b\to c$ decays, it is crucial that
the cut-off parameter $\Delta$ be sufficiently large 
to avoid sensitivity to the shape of the
spectrum in the resonance region. To apply them to 
$b\to u$ decays, $\Delta$ must be small enough to suppress 
the charm background, which sets in 
at $\Delta\sim 650$~MeV. Balancing between the two cases 
restricts the cut-off parameter $\Delta$ to a 
rather small window.  

We also observe that the weight-function analysis with our toy model 
systematically underestimates the result for $F_u(\Delta)$ compared
to the ``true'' result (\ref{eq:Fu}).
For instance, at our reference point $\Delta = 0.65$~GeV,
we have
\beq
  F_u(0.65~{\rm GeV}) &=& 0.79 \qquad \mbox{from (\ref{eq:Fu})} \ , \\
  F_u(0.65~{\rm GeV}) &=& 0.55 \qquad \mbox{from 
toy model and (\ref{eq:weight}), 
      PS scheme} \ .
\label{eq:Futoy}
\eeq
This is due at least in part to the crudeness of our model, 
which completely ignores the non-negligible continuum contribution.  
While we could refine our model to take this into account,
we think that such fine-tuning is best resolved by experimental 
input.

\section{Conclusions}
\label{sec:conclusions}

We analyzed perturbative corrections to 
$\bar B\to X_c \ell \bar\nu_\ell$ decays
using the power counting $m_c\sim \sqrt{\LQCD m_b}$ 
for the charm-quark mass. This treatment implies
that a certain class of partially integrated $b\to c$ decay spectra 
is sensitive to the non-perturbative shape-function 
effects familiar from  $\bar B\to X_u \ell \bar\nu_\ell$ decays.
With the aid of soft-collinear effective theory, we showed that  
the one-loop corrections to such 
decay spectra can be written as a 
convolution of hard, jet, and shape functions.
The hard and shape functions are identical to those
found in the factorization formula for 
$\bar B\to X_u \ell \bar\nu_\ell$ decays,
but the jet function depends explicitly on $m_c$
and hence receives non-trivial corrections
unique to decay into charm quarks.  
We calculated these corrections at NLO
in perturbation theory and at
leading order in the $1/m_b$ expansion, and 
derived a shape-function independent
relation between partially integrated 
$\bar B\to X_c \ell \bar\nu_\ell$ and 
$\bar B\to X_u \ell \bar\nu_\ell$ decay spectra.
This relation can be used to determine $|V_{ub}|$.

Numerical studies raised some issues 
related to this treatment. First,  
the portion of phase-space where the shape-function
approach is valid is somewhat smaller in  
$\bar B\to X_c \ell \bar\nu_\ell$ decays than in the charmless
case. Second, although the results are formally independent 
of the renormalization scheme used to define 
the charm-quark mass,  the numerical  
dependence on the mass scheme is significant. 
Finally, the structure of power
corrections is slightly more complicated than in 
the charmless case, since one encounters not only 
sub-leading shape functions, but also kinematic
power corrections. Some of the power corrections are
enhanced by large logarithms  $\ln(m_c^2/m_b^2)$.

Our study may help improve the understanding
of inclusive $B$ decays in the shape-function region.
On the one hand, it provides additional information
for the extraction of  $|V_{ub}|$. On the other hand,
it may offer an additional testing ground for theoretical
methods based on factorization and soft-collinear effective
theory. To explore these ideas further would require 
experimental information on the partially integrated 
$\bar B\to X_c \ell \bar\nu_\ell$ decay spectrum used
in our analysis.


\section*{Acknowledgements}
This work was supported by the DFG Sonderforschungsbereich SFB/TR09 
``Computational Theoretical Particle Physics'' and by the German
Ministry of Education and Research (BMBF).

\end{document}